\documentclass[twocolumn, amssymb, amsmath, aps, prb, showpacs, 10pt]{revtex4-1}
\usepackage[dvips]{graphicx}
\usepackage[dvips]{color}
\usepackage{xcolor}
\usepackage{hyperref}
\hypersetup{
    colorlinks=true,
    linkcolor=blue,
}

\begin{document}

\title{Remarkable effects of dopant valency $-$ a comparative study of CaBaCo$_{3.96}$Cr$_{0.04}$O$_7$ and CaBaCo$_{3.96}$Ni$_{0.04}$O$_7$}

\author{M. Islam$^1$}
\author{S. Adhikari$^1$}

\author{S. Pramanick$^2$}%
\author{S. Chatterjee$^2$}

 
\author{A. Karmakar$^1$}
 \email{dr.arindam.karmakar@gmail.com}
 \affiliation{$^1$Department of Physics, Surya Sen Mahavidyalaya, Siliguri 734 004, India
}%
 \affiliation{$^2$UGC-DAE Consortium for Scientific Research, Kolkata Centre, Sector III, LB-8, Salt Lake, Kolkata 700 106, India}

\date{\today}

\begin{abstract}
Distinctive effects of dopant valency is discussed using an impurity level 1\% doping each of Cr$^{3+}$ and Ni$^{2+}$ in CaBaCo$_4$O$_7$. Through a comparative study of the magnetic and dielectric properties of multiferroic CaBaCo$_{3.96}$Cr$_{0.04}$O$_7$ and CaBaCo$_{3.96}$Ni$_{0.04}$O$_7$, we highlight that Cr doping does not significantly alter the properties in spite of differences in magnetic spin and ionic radius compared to Co$^{3+}$ and Co$^{2+}$ while Ni doping induces spectacular changes. Particularly, a manifold increase of electric polarization change up to 650 ${\rm \mu}$C/m$^2$ at 5.6 kV/cm is observed in CaBaCo$_{3.96}$Ni$_{0.04}$O$_7$ compared to CaBaCo$_{3.96}$Cr$_{0.04}$O$_7$ and a stronger magnetoelectric coupling leading to a polarization change of $\sim$ 12\% in 15 T magnetic field and below 40 K. Further, magnetodielectric effects hint to competing magnetic phases over the temperature range of magnetic transitions. We discuss the observed disparity in the light of a possible site selective doping of Cr$^{3+}$ and Ni$^{2+}$ in the triangular and kagom\'e layers, respectively, of the parent compound. 
\end{abstract}

\maketitle
\section{Introduction}
At the outset of the present decade a new multiferroic ferrimagnet, CaBaCo$_4$O$_7$, was discovered, showing strong electric polarization~\cite{CBCO-die}. The compound was derived from yet another newly discovered '114' oxide system, $Ln$BaCo$_4$O$_7$ ($Ln$ = Y or Lanthanides) in the last decade~\cite{Val-hex1,Val-hex2,Chap-hex1,Maig-hex,Man-hex1,Caig-hex}. Both of these are characterized by quite distinctive 1:1 alternate stacking of triangular and kagom\'e layers of corner sharing CoO$_4$ tetrahedra~\cite{Mueller} exhibiting geometrical frustration and highly degenerate magnetic ground states. The Co ions are exposed to tetrahedral crystal field which results in a higher energy $t_{2g}$ levels and lower energy $e_g$ levels. Co$^{3+}$ ($d^6$) are arguably in the high spin state while the Co$^{2+}$ ($d^7$) can only be in the high spin state~\cite{Val-hex2}. Additionally, the valence ratio of Co$^{3+}$ and Co$^{2+}$ plays a major role in defining the magnetic state of the materials. These are the key factors that drive the interesting and diverse magnetic properties among the members of the family often induced by even the slightest doping. For example, YBaCo$_4$O$_7$, upon cooling, showed a structural transition at 313 K from trigonal ($P31c$) to orthorhombic ($Pbn2_1$) structure, inheriting distortion, that weakens the frustration leading to an antiferromagnetic (AFM) ordering below 110 K~\cite{Chap-hex1}. Earlier study, however, describes a hexagonal $P6_3mc$ symmetry of YBaCo$_4$O$_7$~\cite{Val-hex2}. Quite differently, CaBaCo$_4$O$_7$ crystallizes in a strongly distorted orthorhombic structure in the polar $Pbn2_1$ space group (reported up to 400 K), bearing 1:1 charge ordering of Co$^{3+}$ and Co$^{2+}$ and shows ferrimagnetic (FIM) ground state below $\sim$ 62 K ($T_N$) of $m^\prime m2^\prime$ symmetry in the magnetic ordering~\cite{Caig,Caig-Neu,XAFS}. Among the four nonequivalent atomic sites of the Co framework, the Co1 alone occupies the triangular layer and the Co2, Co3, Co4 occupy the kagom\'e layer. The latter three sites form nearest neighbor triangular patterns exhibiting geometrical frustration. Co2 and Co3 accommodate Co$^{2+}$ ions while Co1 and Co4 house Co$^{3+}$ ions~\cite{XAFS}. Caignaer {\it et al.} describes the FIM structure of CaBaCo$_4$O$_7$ as consisting of zig-zag chains of Co$^{2+}$ spins located at the Co2 and Co3 sites in the kagom\'e layer running along the $b$-direction. These spins are coupled ferromagnetically along $b$, antiferromagnetically along $a$ and co-planar in the $ab$ plane. Half of the Co$^{3+}$ spins are located in the triangular layer at the Co1 sites and the other half in the kagom\'e layer at the Co4 sites, coupled antiferromagnetically to the Co$^{2+}$ spins and co-planar in the $ab$ plane~\cite{Caig-Neu}. Fascinatingly, the compound also exhibits non-switchable gigantic change in electric polarization ($\Delta P$) along $c$ axis below $T_N$ with a maximum value of $\sim$ 17 mC/m$^2$ at 1.1 kV/cm electric field poling at $\sim$ 7 K (in single crystal study) and strong linear magnetoelectric coupling coefficient of 764 ps/m~\cite{CBCO-single}. The electrical ground state of the compound was argued to be pyroelectric in nature~\cite{John}. The origin of the electric ordering and magnetoelectric coupling has been discussed variously although, in a broader sense, the origin is magnetostriction~\cite{John}. For example, Fishman {\it et al.} attributes the origin of strong polarization in CaBaCo$_4$O$_7$ to the electric field induced modification of a set of bonds connecting three Co bitetrahedral chains running along the $c$ direction~\cite{Fishman} while Dey {\it et al.} demonstrate the improvement of polarization via La doping due to reduced Co interlayer distance along with a two dimensional deformation of the kagom\'e layer~\cite{Giri}.

\par
Owing to the attractive multiferroic properties with large magnetic and electric polarizations, large magnetoelectric coupling and potentials for device applications, this system has been revisited several times over the last decade to understand the origin of multiferroic properties and to find ways to tune the transition temperatures. Quite a few doping studies have been done till date with different observations. Almost all the doping studies using divalent Zn~\cite{Sar-dope2} and trivalent Ga~\cite{Sar-dope2}, Fe~\cite{Motin-dope2} and Al~\cite{Al-dope} have shown a considerable reduction of $T_c$ and saturation magnetization even with impurity level doping fractions. Less than 1\% Zn doping in CaBaCo$_{3.9}$Zn$_{0.1}$O$_7$ induces an AFM state along with a spin-glass phase at $\sim$ 20 K~\cite{Sar-dope1,Sar-dope2}. A spin-glass phase appears in CaBaCo$_{1-x}$Al$_x$O$_7$ above $x=0.1$~\cite{Al-dope} and a frustrated phases separated state in CaBaCo$_{1-x}$Fe$_x$O$_7$ for $x\geq 0.04$~\cite{Motin-dope2}. Further doping generally results in transition to a hexagonal $P6_3mc$ structure. Magnetodielectric effects have been characterized in CaBaCo$_{3.97}$Zn$_{0.03}$O$_7$~\cite{Motin-dope1,Preet}, and the nature of electric ordering upon Ni doping is discussed in another study~\cite{Ni-dope}. In this work we provide a systematic comparison of magnetic and dielectric properties of 1\% trivalent Cr doped CaBaCo$_{3.96}$Cr$_{0.04}$O$_7$ and 1\% divalent Ni doped CaBaCo$_{3.96}$Ni$_{0.04}$O$_7$. We highlight the striking differences of the effects of valency and ionic radii of the dopants. While Cr doping does not have major effects on the properties (contrary to what is observed for 1\% Fe doping $-$ a similar case) in spite of its differences in magnetic spin and ionic size, Ni doping shows remarkable differences. Most significantly, Ni doping leads to manifold enhancement of $\Delta P$ and stronger magnetoelectric coupling. We also discuss the likelihood of site selective entry of dopant ions in the triangular and kagom\'e layers in order to explain the observed properties.

\begin{figure}[t]
\centering
\includegraphics[width=0.8\columnwidth]{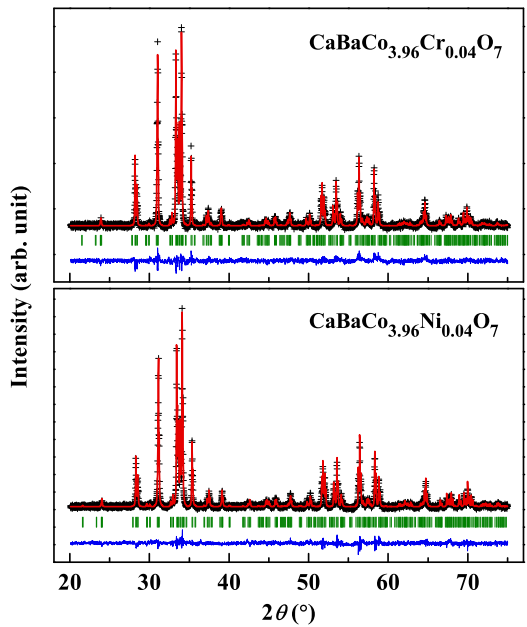}
\caption{Rietveld refinement plots at 300 K. The black symbols and the red solid curves represent the experimental data points and calculated diffraction patterns, respectively. The blue curves are the difference plots and the green markers are the positions of Bragg reflections.}
\label{XRD}
\end{figure}
  
\section{Experiment}
Polycrystalline samples with nominal compositions CaBaCo$_{3.96}$Cr$_{0.04}$O$_7$ (CBCCO) and CaBaCo$_{3.96}$Ni$_{0.04}$O$_7$ (CBCNO) were synthesized by conventional solid state reaction process. Stoichiometric amounts of CaCO$_3$, BaCO$_3$, Co$_3$O$_4$, Cr$_2$O$_3$ and NiO were mixed and then calcined for 12 h at 900$^{\circ}$C in air. The mixtures were then pelletized after thorough mechanical homogenization and heated at 1100$^{\circ}$C in air for 24 h and quenched down to room temperature. The last step was repeated once to improve homogeneity and reduce impurity. Phase identification and structural characterization of prepared samples were done by Bruker D8-Advance X-ray powder diffractometer using Cu-K$_\alpha$ radiation ($\lambda$ = 1.5406 {\AA}) at room temperature. Rietveld refinements were done using FULLPROF suite~\cite{fullprof}. Dielectric measurements were carried out in a 15 tesla commercial cyrogen-free system from Cyrogenic-Ltd, U.K., in the temperature range 5$-$300 K equipped with an Agilent E4980A precision LCR meter and a Keithley 6517A electrometer on thin specimens ($\sim$ 0.4 mm thickness) of parallel plate geometry with large surface area to thickness ratio. Change in polarization ($\Delta P$) was calculated from pyroelectric current measured using the electrometer, considering a zero reference value above the transition temperature. For this, the sample was cooled from 100 K down to 5 K at 1 K/min rate under an applied poling voltage. The sample was then short circuited for 1 h to remove any stray charges, after disconnecting the poling electric field. Pyroelectric current ($I_p$) was measured in the warming cycle at a rate of 4 K/min. Magnetic experiments were done both in the above system with the option of a VSM attachment and also in a Quantum Design MPMS XL system.

\begin{table*}[t]
\centering
\caption{Atomic coordinates for CaBaCo$_{3.96}$Cr$_{0.04}$O$_7$ and CaBaCo$_{3.96}$Ni$_{0.04}$O$_7$ at room temperature (300 K) obtained from Rietveld refinement of X-ray diffraction data using space group $Pbn2_1$. The lattice parameters are CaBaCo$_{3.96}$Cr$_{0.04}$O$_7$: $a=6.2878(9)$ \AA, $b=11.0079(0)$ \AA, and $c=10.1922(4)$ \AA; CaBaCo$_{3.96}$Ni$_{0.04}$O$_7$: $a=6.2908(2)$ \AA, $b=11.0102(6)$ \AA, and $c=10.1929(3)$ \AA. The occupancy of the dopants (Cr and Ni) are 0.02 at each of the sites indicated in the table which makes the Co occupancy 0.98 at the same site.}
\label{Tbl}
\begin{center}
\begin{tabular}{ccccc|ccccc}		
\hline
\hline
Specimen $\rightarrow$ & &&CaBaCo$_{3.96}$Cr$_{0.04}$O$_7$&& &&&CaBaCo$_{3.96}$Ni$_{0.04}$O$_7$&\\
\hline \\
Atom $\downarrow$        & & $x$ &	$y$ &	$z$          &&	Atom $\downarrow$ &	$x$ &	$y$ &	$z$ \\   \\
\hline \\
Ca      &           & -0.0133(8) & 0.6692(4) & 0.8708(6)     && Ca      & 0.0019(8) &  0.6744(7) &  0.8738(2)  \\
Ba      &           & -0.0003(5) & 0.6665(3) & 0.5           && Ba      & 0.0041(1) &  0.6698(2) &  0.5        \\
Co1/Cr1 &           &  0.0123(4) & 0.0034(6) & 0.9425(5)     && Co1     & 0.0067(7) & -0.0142(8) &  0.9451(3)  \\
Co2     &           & -0.0050(7) & 0.1685(9) & 0.6912(1)     && Co2/Ni2 & 0.0204(0) &  0.1662(0) &  0.6831(0)  \\
Co3     &           &  0.2501(4) & 0.0886(5) & 0.1916(7)     && Co3/Ni3 & 0.2472(7) &  0.0905(9) &  0.1867(2)  \\
Co4/Cr4 &           &  0.2718(2) & 0.9125(1) & 0.6861(0)     && Co4     & 0.2504(8) &  0.9133(1) &  0.6879(2)  \\
O1      &           &  0.0453(2) & 0.0096(9) & 0.2428(5)     && O1      & 0.0300(0) & -0.0043(8) &  0.2450(5)  \\
O2      &           &  0.0224(3) & 0.4932(6) & 0.2471(2)     && O2      &-0.0028(7) &  0.4939(5) &  0.2354(3)  \\
O3      &           &  0.8021(4) & 0.2603(4) & 0.7830(2)     && O3      & 0.7700(6) &  0.2392(6) &  0.7670(6)  \\
O4      &           &  0.6637(8) & 0.7601(6) & 0.2092(2)     && O4      & 0.7390(0) &  0.7653(5) &  0.2112(4)  \\
O5      &           & -0.0208(9) & 0.1607(3) & 0.5125(4)     && O5      &-0.0333(8) &  0.1672(8) &  0.5044(5)  \\
O7      &           &  0.2195(3) & 0.1090(0) &-0.0042(3)     && O6      & 0.2142(9) &  0.1080(4) &  0.0025(5)  \\
O8      &           &  0.2751(4) & 0.9445(1) & 0.4913(9)     && O7      & 0.2715(4) &  0.9479(3) &  0.5041(8)  \\
\hline   
\hline
\end{tabular}
\end{center}
\end{table*}     

\vskip 0pt
\section{Results and Discussions}
Rietveld refinement of the X-ray diffraction patterns at 300 K, given in Fig. \ref{XRD}, shows good crystallization of the compound. No impurity reflections are detected within the limitations of the instrument. Dependable $\chi^2$ values (2.15 for CBCCO and 2.36 for CBCNO) and quality of the difference plots substantiate the phase purity of the samples. The atomic positions and the reliability parameters are listed in Table \ref{Tbl}. Both the diffraction patterns are fitted using the orthorhombic $Pbn2_1$ space group similar to the parent compound CaBaCo$_4$O$_7$ (CBCO)~\cite{Caig-Neu}. The lattice parameters obtained are (CBCCO) $a=6.2878(9)$ \AA, $b=11.0079(0)$ \AA, $c=10.1922(4)$ \AA \ and (CBCNO) $a=6.2908(2)$ \AA, $b=11.0102(6)$ \AA, $c=10.1929(3)$ \AA. Refinements in CBCCO, considering Cr ions located at the trivalent Co1 and Co4 sites, produced lower $\chi^2$ value compared to divalent Co2 and Co3 sites for which $\chi^2$ exceeded 3.2. This seems likely because of the stable 3+ oxidation state of Cr. Similarly, refinements in CBCNO produced best results considering Ni doping at the divalent Co2 and Co3 sites, owing to the stable 2+ oxidation state of Ni. These results are in accordance with the previous reports~\cite{Sar-dope2,Motin-dope2,XAFS,Ni-dope}. Some of the compounds in the genre, the $Ln$BaCo$_4$O$_7$~\cite{Maig-hex,Caig-hex,Val-hex1,Chap-hex1,Man-hex1,Val-hex2} and the heavily doped specimens of CBCO~\cite{Sar-dope1,Sar-dope2,XAFS,Motin-dope2}, show a structural transition from a higher symmetry hexagonal $P6_3mc$ to the lower symmetry orthorhombic and polar $Pbn2_1$ space group with decreasing temperature ($T$). The appearance of hexagonal symmetry in case of high doping is found to be due to reduction of orthorhombic distortion~\cite{Sar-dope1,Sar-dope2,XAFS,Motin-dope2,Pra-oxy} quantified by $D=\left(b/\sqrt{3}-a\right)/a$. In our case we found $D=1.08$\% for CBCCO and $1.05$\% for CBCNO at 300 K which is well within the range observed for orthorhombic varieties such as Zn, Ga doping~\cite{Sar-dope2} and Fe doping~\cite{Motin-dope2} in CBCO. The orthorhombic phase is found to sustain up to $D \approx 0.4$\% for Zn doping and 0.5\% for Ga doping~\cite{Sar-dope2}. $D$ for our materials are well within the range for an orthorhombic structure.

\begin{figure}[t]
\centering
\includegraphics[width=1.0\columnwidth]{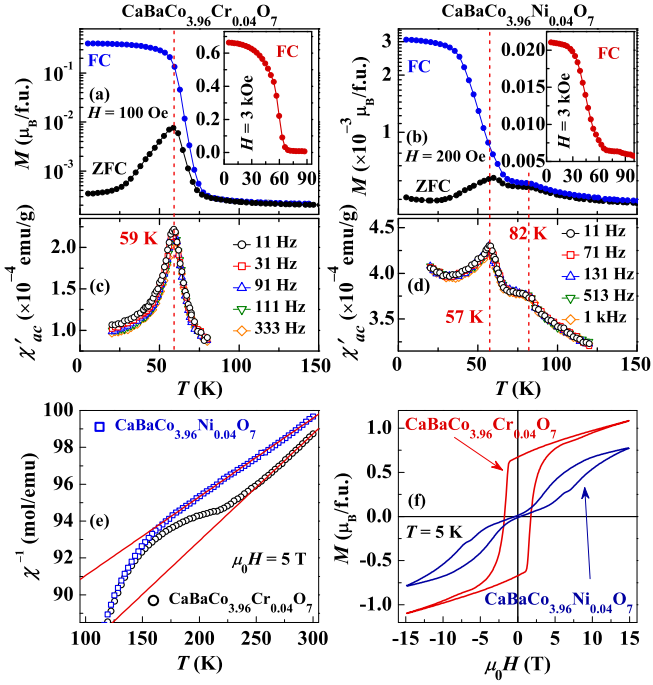}
\caption{(a) and (b) show the magnetization ($M$) with temperature ($T$) of CaBaCo$_{3.96}$Cr$_{0.04}$O$_7$ and CaBaCo$_{3.96}$Ni$_{0.04}$O$_7$, measured in 100 Oe and 200 Oe magnetic fields, respectively. Insets represent the field cooled (FC) magnetization measured in 3 kOe fields. (c) and (d) portray the respective ac susceptibility (real part $-$ $\chi^\prime_{ac}$) with temperature, plotted in the same scale as (a) and (b). $\chi^\prime_{ac}$ is plotted for five measuring frequencies as indicated. The vertical red dashed lines indicate the transition temperatures as a reference for both $M(T)$ and $\chi^\prime_{ac}(T)$ curves. (e) Inverse susceptibility ($\chi^{-1}$) with temperature measured at a high magnetic field of 5 T. The red lines show fits according to Curie-Weiss law. (f) Isothermal magnetization hysteresis loops [$M(H)$] up to 15 T magnetic field recorded at 5 K.}
\label{mag}
\end{figure}

Thermal dependence of dc magnetization [$M(T)$] is shown in Fig. \ref{mag}(a) and (b) measured at low magnetic fields ($H$), as indicated in the figures. The curves are plotted in logarithmic scale in order to highlight the zero field cooled (ZFC) peak which is otherwise suppressed by the strong field cooled (FC) branch. Similar to CBCO~\cite{Caig} and other low doped compounds~\cite{Motin-dope2,Sar-dope2}, the $M(T)$ curves for both the samples show bifurcation quite above the sharp ZFC peak indicating frustration in magnetic ordering. Thermal variations of the real part of ac susceptibility [$\chi^\prime_{ac}(T)$] are plotted in Fig. \ref{mag}(c) and (d) for CBCCO and CBCNO, respectively. The order of magnitudes observed are consistent with other analogous materials~\cite{Caig,Ni-dope,Motin-dope2,Sar-dope2}. $\chi^\prime_{ac}(T)$ of CBCCO shows a sharp peak at 59 K nearly coinciding with the ZFC-$M(T)$ peak in the dc magnetization. CBCNO also shows a similar sharp peak in $\chi^\prime_{ac}(T)$ at 57 K coinciding with the dc ZFC-$M(T)$ peak along with another step-like feature at 82 K, also observed in dc ZFC-$M(T)$. Similar kink in $\chi^\prime_{ac}(T)$ was observed around the region of 80 K in other studies although there is no detailed picture till date about the magnetic ordering~\cite{Ni-dope,Motin-dope1,Motin-dope2}. For example, it was evident in 0.75$-$10\% Zn doping (divalent) which was ascribed to a new weak AFM ordering~\cite{Sar-dope2} and in Ni-doping (divalent) ascribed to short-range interactions due to absence of magnetic reflections above 65 K in neutron diffraction patterns~\cite{Ni-dope}. We denote the transition by $T_s$. We also find from previous studies that a weaker but distinct kink is observed in Ga doping (trivalent)~\cite{Sar-dope2} at least up to 3.75\% and also in trivalent Fe doping in the range 0.25$-$1.25\%~\cite{Motin-dope2}. Interestingly, our 1\% trivalent Cr ($3d^3-$high spin state 3/2$\mu_B$) doped sample does not show such kink, a behavior similar to trivalent nonmagnetic Al ($2p^6$) doping~\cite{Al-dope}, but the 1\% divalent Ni ($3d^8-$spin 1$\mu_B$) doped sample does. The behavior seems to be related to the size matching of the dopant ions involved, the size of which dictates the target doping site in the parent material, being either in the triangular layer or in the kagom\'e layer. Table \ref{Tbl3} summarizes the ionic radii of the dopants and doping site ions. (As the ionic radii of Co$^{3+}$ and Cr$^{3+}$ in tetrahedral coordination are not found, we have included a possible range of the values.) It may be noted that radii of Zn$^{2+}$ and Ni$^{2+}$ are close to that of Co$^{2+}$ at the Co2 and Co3 atomic sites in the kagom\'e layer and can be easily accommodated there. Ga$^{3+}$ and Fe$^{3+}$ being a bit larger than or comparable to Co$^{3+}$, these ions preferentially enter the trivalent Co4 site in the kagom\'e layer rather than trivalent Co1 sites in the triangular layer. This is because as the kagom\'e layer already accommodates larger Co$^{2+}$ ions, the size constraint is more flexible in it than the triangular layer which accommodates only the smaller Co$^{3+}$ ions. Al$^{3+}$ and Cr$^{3+}$ however are way smaller than Co$^{2+}$. These ions being even smaller than or similar to Co$^{3+}$ may be preferentially doped at the Co1 atomic site in the triangular layer in low doping cases to save the cost of elastic deformation of the surrounding atoms on being located at the kagom\'e layer. It thus seems that the appearance of the kink around 80 K is related to doping in the kagom\'e layer while it is absent for doping in the triangular layer. It seems plausible because the magnetic ground state is mainly governed by the ordering in the kagom\'e layer. It may also be noted that the step at $T_s$ in $\chi^\prime_{ac}(T)$ in CBCNO does not show any shift with frequency. A short range ordering is thus unlikely at $T_s$ but a weak long-range magnetic ordering seems to be the origin.
\begin{table}[t]
\centering
\caption{Ionic radii in a crystal solid in tetrahedral coordination obtained from the work of R. D. Shannon~\cite{Shannon}. "HS" in parenthesis indicates high spin state.}
\label{Tbl3}
\begin{center}
\resizebox{0.6\columnwidth}{!}{
\begin{tabular}{ccccc}		
\hline
Ion       & Radius (pm) && Ion       & Radius (pm)\\
\hline \\
Co$^{2+}$ & 72 (HS)     && Co$^{3+}$ & 54$-$60 \\
Zn$^{2+}$ & 74          && Ni$^{2+}$ & 69 \\
Ga$^{3+}$ & 61          && Fe$^{3+}$ & 63 (HS) \\
Al$^{3+}$ & 53          && Cr$^{3+}$ & 55$-$61 \\
\hline
\end{tabular}
}
\end{center}
\end{table}     
\begin{table}[t]
\centering
\caption{Transition temperatures and magnetizations reported for similar compounds.}
\label{Tbl2}
\begin{center}
\resizebox{\columnwidth}{!}{
\begin{tabular}{ccccccccccc}		
\hline
Compound                          && $T_c$(K) && $M$($\mu_B$/f.u.) && @ $T$(K) && @ $H$(Oe) && Ref\\
\hline \\
CaBaCo$_{3.96}$Cr$_{0.04}$O$_7$   && 59       && 0.66 					  && 5        && 3k        && Our work \\
CaBaCo$_{3.96}$Ni$_{0.04}$O$_7$   && 57       && 0.021					  && 5        && 3k        && Our work \\
CaBaCo$_4$O$_7$ (polycrystal)     && 60       && 0.67						  && 5        && 3k        && \onlinecite{Caig-Neu,Motin-dope2} \\
CaBaCo$_4$O$_7$ (single crystal)  && 64       && ---						  && --       && --        && \onlinecite{CBCO-single} \\
CaBaCo$_{3.96}$Fe$_{0.04}$O$_7$   && 42       && 0.028					  && 5        && 3k        && \onlinecite{Motin-dope2} \\
CaBaCo$_{3.95}$Ga$_{0.05}$O$_7$   && 40       && 0.021					  && 5        && 3k        && \onlinecite{Sar-dope2} \\
CaBaCo$_{3.96}$Ni$_{0.04}$O$_7$   && 60       && 0.0195					  && 2        && 3k        && \onlinecite{Ni-dope} \\
CaBaCo$_{3.96}$Zn$_{0.04}$O$_7$   && 40       && 0.03					    && 2        && 3k        && \onlinecite{Sar-dope2} \\
CaBaCo$_{3.98}$Al$_{0.02}$O$_7$   && 52       && 0.60				      && 2        && 1k        && \onlinecite{Al-dope} \\
CaBaCo$_{3.95}$Al$_{0.05}$O$_7$   && 32       && 0.0064				    && 2        && 1k        && \onlinecite{Al-dope} \\

\hline
\end{tabular}
}
\end{center}
\end{table}     

\par
To sort out the nature of the main transition in both samples, it is pertinent to point out the following points. (I) Neither of the features observed in $\chi^\prime_{ac}(T)$ in either sample, particularly the sharp peaks, show any discernible peak shift with increasing frequency indicating a long-range magnetic ordering of AFM origin. (II) The irreversibility between the dc ZFC and FC-$M(T)$. (III) A step-like ferromagnetic nature of FC--$M(T)$. All these ensure FIM nature of the transitions [indicated by the $\chi^\prime_{ac}(T)$ peaks] with an inbuilt disorder inherent in the geometrical frustration, similar to the ground state of CBCO and other low doped materials~\cite{Caig,Caig-Neu,Motin-dope2,Ni-dope,XAFS,Sar-dope2}. The peak temperatures of $\chi^\prime_{ac}(T)$ are thus considered as the FIM transition temperatures ($T_c$). $T_c=59$ K for CBCCO and $T_c=57$ K for CBCNO. $T_c$ of CBCCO nearly matches with that of polycrystalline CBCO but is a bit lower than that reported for single crystal (see Table \ref{Tbl2}) while that of CBCNO is a bit lower than polycrystalline CBCO although another study found that it remains same ($\sim$ 60 K)~\cite{Ni-dope}. Table \ref{Tbl2} provides a comparison of the values of $T_c$ and FC magnetization with other compounds of similar impurity level doping at the Co-site, obtained from previous reports. One may note that the magnitude of the FC magnetization at 3 kOe of CBCCO [inset of Fig. \ref{mag}(a)] remains almost same as CBCO but it is markedly reduced in case of CBCNO [inset of Fig. \ref{mag}(b)] by a factor of $\sim$ 31. Quite contrary to our observation, other doping studies such as trivalent Fe and Ga doping showed a marked reduction in $T_c$ by $\sim$ 20 K while Al doping showed a reduction by $\sim$ 30 K. The FC magnetizations of Fe and Ga doped samples are lower by an order of magnitude. Doping study using divalent Zn also shows a reduced $T_c$ by 20 K while the order of FC magnetization remains same as compared to our case of divalent Ni doping.

\par
Fig. \ref{mag}(e) shows the inverse susceptibility plots [$\chi^{-1}(T)$] up to 300 K, fitted with Curie-Weiss law $\chi^{-1}=C/(T-\theta_{CW})$.
$\chi^{-1}$ of CBCCO starts to deviate from linearity below $\sim$ 260 K, moving above the linear fit, signifying short range magnetic interactions. Large negative value of $\theta_{CW}$ obtained (= $-$1402 K), indicating strong AFM interactions, is as per other studies~\cite{Caig,Pra-oxy,Motin-dope2}. CBCNO however shows a usual behavior but a larger negative $\theta_{CW}$ (= $-$2037 K) is obtained. Considering only the possibilities of high-spin state as described by Valdor~\cite{Val-hex2}, and the doping percentages, the effective paramagnetic moment ($p_{eff}$) is expected to be 13.96 for both. For CBCCO we obtain a value of 11.74 from the $\chi^{-1}(T)$ plot while for CBCNO we obtain 13.70. Lower value in CBCCO points to a higher degree of geometrical frustration while a value close to the upper limit in case of CBCNO hints that Ni doping has relieved the frustration, possibly by structural modifications. Isothermal magnetic hysteresis [$M(H)$] loops at 5 K are shown in Fig. \ref{mag}(f) up to 15 T. Neither of the loops close even at 15 T. The $M(H)$ loop of CBCCO shows a large ferromagnetic like opening in the low field region and an AFM like increasing trend at high fields reminiscent of ferrimagnetic ground state with a large coercive field ($H_c=17.6$ kOe), large remanent magnetization ($M_r=0.67$ $\mu_B$/f.u.) and a large magnetization at 15 T ($M_{\rm max}=1.08$ $\mu_B$/f.u.). The $M_{\rm max}$ is fairly small compared to a FM state, thus indicating a FIM ground state of the compound. The values obtained are comparable to CBCO and are close to 0.5\% Al doping~\cite{Al-dope}. It may be noted that both 1.25\% Al doping~\cite{Al-dope} and 1\% Fe doping~\cite{Motin-dope2} destroy the FIM state in the parent compound but it is sustained in Cr doping up to 1\% and to some extent in 1\% Ga doping~\cite{Sar-dope2}. CBCNO shows a distinct two loop nature of the $M(H)$ curve indicating a diffused spin-flop transition with an overall FIM nature which can be compared to the previous report on Ni-doping showing $M(H)$ up to 5 T field~\cite{Ni-dope}. The spin-flop transition typically indicates weak anisotropy energy in the material. A hint of the metamagnetic transitions can also be observed in the $M(H)$ loops of 1\% Zn doped~\cite{Sar-dope2} and 1\% Fe doped~\cite{Motin-dope2} compounds but not robust like 1\% Ni doping. The loop does not close until 15 T field in CBCNO too and shows a low $H_c$ (= 3.7 kOe), low $M_r$ (= 0.011 $\mu_B$/f.u.) and low $M_{\rm max}$ (= 0.78 $\mu_B$/f.u.).

\begin{figure}[t]
\centering
\includegraphics[width=1\columnwidth]{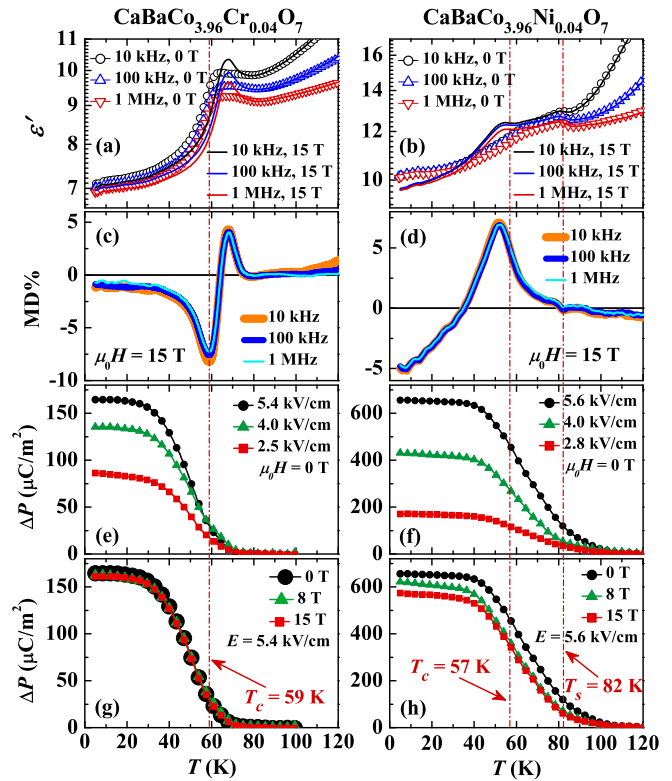}
\caption{(a) and (b) show the temperature dependence of the real part of dielectric permittivity ($\epsilon^\prime$) in 0 and 15 T magnetic field, respectively, for three signal frequencies. (c) and (d) portray the temperature dependence of magnetodielectric percent (MD\%) calculated from the data in (a) and (b), respectively. (e) and (f) display the change in electric polarization ($\Delta P$) as a function of temperature at three different poling electric fields, in zero magnetic field. (g) and (h) show $\Delta P$ as a function of temperature measured in 0, 8 and 15 T magnetic fields. The vertical red dashed lines indicate the magnetic transition temperatures.}
\label{die}
\end{figure}

\par
Thermal dependence of the real part of dielectric permittivity [$\epsilon^\prime(T)$] is shown in Fig. \ref{die} (a) and (b), with and without magnetic field. The zero field data show a step-like feature in both the samples, close to $T_c$. The inflection points on the low-$T$ sides of the steps coincide with $T_c$. Another broad peak is observable in case of CBCNO, coinciding with $T_s$. The absence of frequency ($f$) dispersion of the features indicates long range electric order. As discussed by Johnson {\it et al.}, the nature of the electric ordering is pyroelectric~\cite{John}. It may be noted that although we observed switching of polarization in negative electric field in both the samples (not shown), the switching in case of a polycrystalline specimen is not conclusive of a ferroelectric nature, hence the discussion is excluded. Application of 15 T magnetic field converts the step-like feature to a peak at the pyroelectric transition. Similar evolution of the step to peak is also observed in CaBaCo$_{3.96}$Zn$_{0.04}$O$_7$~\cite{Motin-dope1}. CBCO however shows the opposite; decreasing sharpness of the peak upon application of magnetic field~\cite{CBCO-die}. Also, the peak is shifted towards higher $T$ in CBCCO and towards lower $T$ in CBCNO. The peak at $T_s$ in CBCNO, however, is not displaced in magnetic field. The concomitant magnetic and electric transitions and the $H$-dependent evolution of the peak near $T_c$ signifies magnetoelectric coupling. It may be noted that the transition at $T_s$ in CBCNO also shows magnetoelectric coupling. The percent of magnetodielectric effect [MD\% = $\{\epsilon^\prime(H)-\epsilon^\prime(0)\}/\epsilon^\prime(0)\times 100$] measured at $\mu_0H=15$ T is plotted in Fig. \ref{die} (c) and (d). Quite large magnitude of |MD\%| $\sim$ 8.3 is observed in CBCCO and 6.9 in CBCNO. The maximum value is observed in the temperature range around the FIM transition, a behavior also seen in other compounds of the family~\cite{Motin-dope1,CBCO-die}. Interestingly, MD\% nearly vanishes at $T_s$ in CBCNO and small negative values above $T_s$ may be due to extrinsic effects as it can be seen form Fig. \ref{die}(b) that $\epsilon^\prime(T)$ shows $f$-dependent dispersion in this $T$-range. Notably, MD\% in CBCCO begins to have non-zero value from 80 K and then switches from positive to negative value close to $T_c$ as $T$ decreases following a peak in the positive side and also followed by a peak in the negative side. But in CBCNO the data mostly remains positive over a large $T$-range covering both the magnetic transitions and particularly starting from $T_s$ as $T$ decreases and switches to negative values at 35 K, fairly below $T_c$, following a peak at 52 K. The behavior suggests contributions from two magnetic phases $-$ (1) a high-$T$ phase just below $T_s$ in CBCNO, and in the range $T_c<T<$80 K in CBCCO that generates positive contribution to MD and (2) the gradually appearing low-$T$ FIM phase (as $T$ decreases) generating negative contribution. The data observed is the net effect of the two. A two phase scenario was also speculated in CaBaCo$_{3.96}$Zn$_{0.04}$O$_7$~\cite{Motin-dope1}. Temperature dependent neutron diffraction experiment is required to elucidate this point.

\par
Change in electric polarization ($\Delta P$) is shown with temperature in Fig. \ref{die} (e) and (f) in zero magnetic field for three poling electric fields ($E$). The most significant observation of this work is that the low-$T$ saturation value of $\Delta P$ is apparently reduced in CBCCO (164 ${\rm \mu}$C/m$^2$ at 5.4 kV/cm) compared to CBCO~\cite{CBCO-die} while it is largely enhanced in CBCNO reaching a value of $\sim$ 650 ${\rm \mu}$C/m$^2$ at 5.6 kV/cm, a nearly four fold enhancement. The increase of $\Delta P$ for increasing $E$ from 4 to $\approx$ 5.5 kV/cm is much larger in CBCNO compared to that in CBCCO indicating a possible saturation of $P(E)$ loops in CBCCO at a value of $E$ in the vicinity of 5.5 kV/cm but much above that value in case of CBCNO. Apart from this, the up-rise of $\Delta P(T)$ around $T_c$ is quite sharp in CBCCO (similar to CBCO~\cite{CBCO-die}) while that of CBCNO occurs over a larger $T$-range covering both $T_c$ and $T_s$. Notably, in CBCNO, the short range magnetic ordering below $T_s$ also gives rise to polarization, as $\Delta P(T)$ starts to increase around $T_s$. $\Delta P(T)$ measured in 8 and 15 T magnetic field is represented in Fig. \ref{die} (g) and (h). Application of magnetic field reduces the value of $\Delta P(T)$ which is consistent with the negative MD effect observed at low-$T$, as discussed above. CBCCO shows very small change with $H$ $\sim$ 2\% at 5 K and 15 T. CBCNO however shows a considerable change of $\sim$ 12\% below 40 K and 15 T resembling a stronger magnetoelectric coupling in the later in the concerned temperature region. As the materials are magnetoelectric, the distinct behavior of the polarization ought to have its origin in the magnetic structure and magnetic ordering. It thus appears that doping at the Co1 site in the triangular layer has a detrimental effect on $\Delta P$ but doping at the kagom\'e layer enhances it, via magnetic ordering. Similar enhancement of polarization (nearly 300 ${\rm \mu}$C/m$^2$ at 0.85 kV/cm) was also observed in Zn$^{2+}$ doping in the kagom\'e layer in CaBaCo$_{3.97}$Zn$_{0.03}$O$_7$~\cite{Preet}.

\begin{figure}[t]
\centering
\includegraphics[width=1\columnwidth]{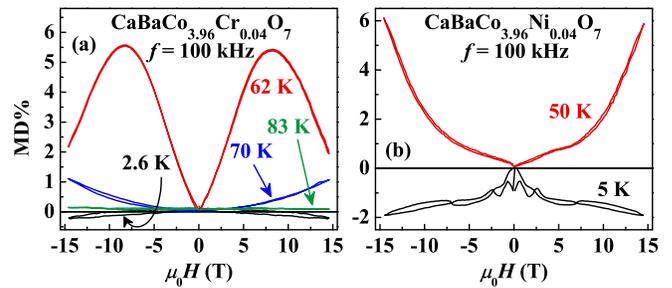}
\caption{Magnetic field dependence of isothermal magnetodielectric percent (MD\%) measured at different temperatures, as indicated.}
\label{md}
\end{figure}

Isothermal magnetodielectric effect as a function of magnetic field [MD\%($H$)] is shown in Fig. \ref{md}. We mainly measured these data at the lowest temperature and in the vicinity of $T_c$. The magnitude of MD\% is very small $\sim$ 0.2 at 2.6 K (CBCCO) and increases to 5.6 at 62 K, close to $T_c$, and decreases drastically above $T_c$ reaching a maximum of 1 at 70 K and 0.1 at 83 K. Thus MD\% shows large values in the region close to $T_c$ only. Notably, the data shows positive values well above $T_c$ and negative values below. In between (at 62 K) it shows a peak-like maximum on either side, a behavior similar to that observed in CBCO~\cite{CBCO-die} and 1\% Zn doping~\cite{Motin-dope1}, which was attributed to the balance between responses from two unique magnetic states, the FIM ground state giving rise to negative MD effect and another magnetic field induced state in the intermediate $T$-region giving rise to positive MD effect (also discussed before). Similar opposite MD effect is also observed in CBCNO. A complete negative curve is observed at 5 K and positive at 50 K. It may be noted with reference to Fig. \ref{die}(d) that the transition from positive to negative MD effect occurs at a much lower $T$ below $T_c$ in CBCNO. The magnitude is again low (1.9\%) at 5 K and higher (6\%) at 50 K, close to $T_c$. 
\section{Conclusions}
This study not only points out the significant differences in the outcome of impurity level dopings of either trivalent or divalent 3$d$ transition metals at the Co-site of CaBaCo$_4$O$_7$ but also points out the unique magnetism and dielectric properties as a result of Cr and Ni doping, which is markedly different from other 3$d$ dopant studies of similar doping percent. While the thermal and field dependent magnetism of CBCCO are similar to CBCO in spite of the differences in the ionic radius and spin moment of Cr$^{3+}$ with Co$^{3+}$ and Co$^{2+}$, CBCNO shows striking disparity both in the nature and magnitude owing to the difference in the spin state alone, the ionic radius of Ni$^{2+}$ being similar to Co$^{2+}$. We propose that Cr$^{3+}$ is preferentially doped at the trivalent Co1 atomic sites in the triangular layer (rather than even the trivalent Co4 sites in the kagom\'e layer) which consequently does not alter the robust ferrimagnetic structure of the parent compound, being governed mainly by the ordering in the kagom\'e layer. This consistently explains the results that the $T_c$, magnitude of FC magnetization, nature of $M(T)$ and $M(H)$ and the parameters obtained from $M(H)$ are similar to the parent compound. Ni$^{2+}$ ions are however doped at the divalent Co2 and Co3 sites in the kagom\'e layer which is expected to considerably disrupt the magnetic ordering. This is evident from the lower magnitude of FC magnetization, somewhat reduced $T_c$, an additional magnetic ordering at $T_s$, completely different nature of $M(H)$ curve with an additional spin-flop transition and radically reduced remanent magnetization, coercive field and 15 T magnetization. Additionally, inverse susceptibility indicates that Ni doping considerably releases the geometrical frustration at 5 T magnetic field but the Cr doping does not. This observation also supports our proposition that Cr ions preferably enter the triangular layer which does not affect the nearest neighbor triangular lattice configuration of Co in the kagom\'e layer mainly responsible for the geometrical frustration. XAFS study is necessary to establish our proposition.

\par
The study also reveals the striking disparity in the dielectric properties and electric polarization which is a consequence of the diverse magnetism of the (magnetoelectric) materials. The temperature dependence of MD\% shows inversion to negative values at much lower temperature ($\sim$ 35 K) in CBCNO compared to $\sim$ 65 K (close to $T_c$) in CBCCO indicating competing magnetic phases at the transition that extend far below $T_c$ in CBCNO. Interestingly, the additional weak magnetic ordering at $T_s$ also exhibits magnetoelectric coupling. The most significant observation is the large difference in the magnitude of electric polarization $\Delta P$. While CBCCO shows a magnitude of 164 ${\rm \mu}$C/m$^2$ at 5.4 kV/cm, seemingly lower than CBCO, CBCNO shows a far higher value of 650 ${\rm \mu}$C/m$^2$ at 5.6 kV/cm. The slight lowering of the value in CBCCO and manifold increase in CBCNO is ascribed to the modification of the magnetic structure due to doping in the triangular layer and kagom\'e layer, respectively. Additionally, the influence of magnetic field on $\Delta P$ is also pronounced in CBCNO and very weak in CBCCO. We suppose that magnetostriction is the origin of the electric polarization in the materials, as described by Johnson {\it et al.}~\cite{John}, and as substantiated by our observations $-$ peak in $\epsilon^\prime(T)$ at $T_c$ and $T_s$, peaks in |MD\%| close to $T_c$, appearance of $\Delta P$ at the magnetic transitions and magnetic field dependence of $\Delta P$. We propose low temperature synchrotron diffraction studies in order to directly probe the effects of magnetostriction and  elucidate the origin of electric polarization.
 
\vskip 2cm
\section{Acknowledgment}
This work was performed under the Collaborative Research Scheme No. UGC-DAE-CSR-KC/CRS/19/MS05/0936 of UGC-DAE Consortium for Scientific Research, Kolkata Center. A. Karmakar and M. Islam acknowledges the grant of the R\&D project. The authors would like to thank the authorities of the Center for providing infrastructural support for materials preparation, characterization, magnetic and dielectric experiments and particularly Dr. P. V. Rajesh for technical support in X-ray powder diffraction measurements.


%

\end{document}